\documentclass[twocolumn,english,preprintnumbers,amsmath,amssymb,floatfix]{revtex4}

\usepackage{epsfig}
\usepackage{graphics}
\usepackage{latexsym}
\usepackage{amsmath}
\usepackage{amssymb}
\usepackage{rotating}
\usepackage{subfigure}
\usepackage{bm}
\usepackage{color}
\usepackage{blindtext}
\usepackage[T1]{fontenc}
\usepackage{blindtext}
\usepackage[latin9]{inputenc}
\usepackage{color}
\usepackage{array}
\usepackage{amstext}
\usepackage{graphicx}
\usepackage{esint}
\usepackage{float}
\makeatletter

\usepackage[colorlinks = true,
linkcolor = magenta,
urlcolor  = blue,
citecolor = red,
anchorcolor = blue]{hyperref}

\@ifundefined{textcolor}{}
{%
	\definecolor{BLACK}{gray}{0}
	\definecolor{WHITE}{gray}{1}
	\definecolor{RED}{rgb}{1,0,0}
	\definecolor{GREEN}{rgb}{0,1,0}
	\definecolor{BLUE}{rgb}{0,0,1}
	\definecolor{CYAN}{cmyk}{1,0,0,0}
	\definecolor{MAGENTA}{cmyk}{0,1,0,0}
	\definecolor{YELLOW}{cmyk}{0,0,1,0}
}

\@ifundefined{definecolor}
{\usepackage{color}}{}
\@ifundefined{definecolor}
{\usepackage{color}}{}
\makeatother

\makeatother

\usepackage{babel}

\begin{document}
	
	\title{ $\Lambda_c^+$ fragmentation functions from pQCD approach and the Suzuki model}

	\author{Mahdi Delpasand$^{a}$}
    \email{Delpasand.mahdi@yahoo.com}

	\author{S.~Mohammad Moosavi Nejad$^{a,b}$}
	\email{mmoosavi@yazd.ac.ir}
	
	\author{Maryam Soleymaninia$^{b}$}
	\email{maryam_soleymaninia@ipm.ir}
	
	\affiliation{$^{(a)}$Faculty of Physics, Yazd University, P.O. Box
		89195-741, Yazd, Iran}
	
	\affiliation{$^{(b)}$School of Particles and Accelerators,
		Institute for Research in Fundamental Sciences (IPM), P.O.Box
		19395-5531, Tehran, Iran}

	\date{\today}

%
\begin{abstract}

Through data analysis, we present new sets of nonperturbative fragmentation functions (FFs) for  $\Lambda_c^+$ baryon both at leading and  next-to-leading order (NLO) and, for the first time, at next-to-next-to-leading order (NNLO) in the minimal subtraction factorization scheme  with five massless quarks. 
The FFs are determined by fitting all available data of inclusive single $\Lambda_c^+$ baryon production in $e^+e^-$ annihilation taken by the {\tt OPAL} Collaboration at CERN LEP1 and {\tt Belle} Collaboration at KEKB.
We  also estimate the uncertainties in the $\Lambda_c^+$ FFs as well as in the corresponding observables. In a completely different approach based on the Suzuki model, we will theoretically calculate the  $\Lambda_c^+$ FF  from charm quark and present our result at leading order perturbative QCD framework. A comparison confirms  a good consistency between both approaches. We will also apply the  $\Lambda_c^+$ FFs to make   theoretical predictions for the energy distribution of $\Lambda_c^+$  produced through the top quark decay, to be measured at the CERN LHC.

\end{abstract}
%

             
\maketitle

\section{Introduction}\label{sec:introduction}

Study of heavy hadrons properties provides a possibility  for better understanding the quark-gluon interaction dynamics in the QCD framework. 
In this regards and due to ongoing experiments there are particular interests in hadron productions at the CERN LHC and the BNL Relativistic Heavy Ion Collider (RHIC). In this work, we study the production mechanism of heavy baryons through the fragmentation process. 
In a general expression, the fragmentation mechanism describes the hadronization process where a parton carrying large transverse momentum decays to form a jet containing the expected hadron \cite{Braaten:1993rw}. The hadronization processes are described by the fragmentation functions (FFs) which refer to the probability densities of hadron productions from  initial partons and make the nonperturbative aspects of hadroproduction processes. These functions along with the parton distribution functions (PDFs) \cite{Salajegheh:2018hfs} construct nonperturbative inputs for the calculation of hadroproduction cross sections. In this work, we focus on the FFs of  $\Lambda_c^+$ baryon through two different approaches. In the first approach which is usually called as the phenomenological approach, a specific form including some free parameters is proposed for the desired FF so that  all these free parameters are extracted through experimental data analysis, see, for example, Refs.~\cite{Kniehl:2012ti,Binnewies:1998vm,Kniehl:2008zza,Salajegheh:2019srg,Nejad:2015fdh,Soleymaninia:2017xhc,Salajegheh:2019ach,Salajegheh:2019nea,Soleymaninia:2018uiv,Soleymaninia:2019sjo,Mohamaditabar:2018ffo,Anderle:2015lqa,Bertone:2017tyb}. 
In an alternative scheme based on the theoretical models, the heavy hadron FFs might be computed by the virtue of perturbative QCD with limited phenomenological 
parameters, see, for example, Refs.~\cite{GomshiNobary:1994eq,MoosaviNejad:2017rvi,MoosaviNejad:2019,Nejad:2013vsa,MoosaviNejad:2018ece,MohammadMoosaviNejad:2018pyp,MoosaviNejad:2016qdx,Nejad:2014iba} where the heavy hadron  FFs are computed by use of the Suzuki model \cite{Suzuki:1977km}. This elaborate model is related to the perturbative QCD framework where all convenient Feynman diagrams at each order of perturbative QCD are considered for the parton level of hadronization process. In this approach, the nonperturbative aspect of hadronization is emerged in the bound state wave function. Detailed description is presented in  Section~\ref{sec:theoretical}.\\
Independent of approach used to extract the FFs, when these functions are computed at the initial scale of fragmentation they can be evolved to higher scales using the timelike
Dokshitzer-Gribov-Lipatov-Altarelli-Parisi (DGLAP) evaluation equations \cite{DGLAP}. \\
In this paper, we first use the  phenomenological approach to obtain a set of gluon,  charm- and bottom-quark FFs  into the $\Lambda_c^+$ baryon  through a global QCD fit to all available $e^+e^-$ single inclusive annihilation (SIA) data measured by {\tt OPAL} Collaboration~\cite{Alexander:1996wy,Ackerstaff:1997ki} at the CERN LEP1 and  very recent data from {\tt Belle} Collaboration \cite{Niiyama:2017wpp}. In Ref.~\cite{Kniehl:2006mw}, the $\Lambda_c^+$  FFs were determined both at leading and next-to-leading order in the minimal subtraction factorization scheme ($\overline{MS}$) by fitting the fractional energy spectra of $\Lambda_c^+$ baryon measured by the {\tt OPAL} in the $e^+e^-$ annihilation on the Z-boson resonance. 
In their work, authors have applied the zero-mass variable-flavor-number scheme (ZM-VFNS) or, in a short expression, the massless scheme, where heavy quarks are treated as massless partons as well.
Recently, {KKSS20} Collaboration has updated their previous analysis \cite{Kniehl:2006mw} combining the data for $e^+e^-$ annihilation from  {\tt OPAL} and the recent one from {\tt Belle}, see Ref.~\cite{Kniehl:2020szu}. Their strategy for constructing the $\Lambda _c^+$-FFs  is the same as their previous work for D-meson  \cite{Kneesch:2007ey}. Their work is restricted to the NLO accuracy and they have not also evaluated  the uncertainties  for FFs and corresponding observables.\\
In the present work,  we focus on the hadronization of gluon,  charm- and bottom-quarks into the $\Lambda_c^+$  using the massless scheme and provide the first QCD analysis of $(g, c, b) \to \Lambda_c^+$ FFs at next-to-next-to-leading order (NNLO). Meanwhile, we go beyond Refs.~\cite{Kniehl:2006mw,Kniehl:2020szu} and perform a full-fledged error estimation   for the parton FFs as well as the resulting
differential cross sections. In order to evaluate the error estimations we apply the well-known  Hessian approach \cite{Pumplin:2000vx}. 
Note that,  our  analysis is restricted to only three data sets from single inclusive annihilation process due to two  reasons: firstly, we are not aware of any other such data from electron-positron annihilation, and secondly, due to the lack of other theoretical partonic cross sections for the production of partons at NNLO accuracy.  Although,  among all processes producing baryons, the $e^+e^-$ annihilation process provides the cleanest environment to calculate the FFs, being devoid of nonperturbative effects beyond fragmentation itself.
\\
In the following and in a theoretical approach independent of data analysis,  we compute the initial scale fragmentation function of charm quark to split into the S-wave $\Lambda_c^+$ baryon  at lowest-order of perturbative QCD. For this approach, we employ the elaborate Suzuki model which contains most of kinematical and dynamical aspects of hadroproduction  process. Finally, we shall compare the initial scale FF of $c\to \Lambda_c^+$ determined in both approaches. Our comparison shows a good consistency between both results. 

In the Standard Model (SM) of particle physics, the top quark has very short life time so does not have enough time to form a bound state, then before it decays hadronization takes place. At the lowest order of perturbative QCD and at the parton level, the decay mode $t \to bW^+$ followed by $b\to X+$Jets is governed. Here, X refers to the detected hadrons in the final state. Thus, at the CERN LHC a proposed channel to indirect search for the top quark properties is to study the energy spectrum of produced hadrons through top decays. In this work, as an example of possible applications of extracted FFs, we make the theoretical predictions for the energy distributions of $\Lambda_c^+$ baryons in  top quark decays at LO, NLO and NNLO. This prediction will be compared with the one obtained through using the fit parameters reported in Ref.~\cite{Kniehl:2020szu}. This comparison does also shows a good consistency between our analysis and the one performed in  Ref.~\cite{Kniehl:2020szu}.

The outline of this paper is as follows:
In Section~\ref{sec:phenomeno framework},  we explain the theoretical framework of hadron production in $e^+e^-$ annihilation  in
the massless scheme and introduce our parametrization of the  $c/b\to \Lambda_c^+$ FF at
the initial scale. We will also describe the minimization method in
our analysis and the approach used for determination of  error estimation.  All applied experimental data will be 
describe in this section and our LO, NLO and NNLO results will be presented and
compared with the data fitted to.  In Section~\ref{sec:theoretical},  through the perturbative QCD approach we provide a general discussion of the fragmentation process for the {\it{S}}-wave heavy baryon and determine the fragmentation distribution of c-quark to fragment into $\Lambda_c^+$ baryon at lowest-order of perturbative QCD.
In Section.~\ref{sec:B-meson-LHC},  predictions for the
normalized-energy distributions of $\Lambda_c^+$ baryon produced from 
top decay are presented. Our conclusions are listed in Section~\ref{sec:conclusion}.

\section{Phenomenological approach  for determination of  $\Lambda_c^+$ FFs and their uncertainties} \label{sec:phenomeno framework}

As was mentioned, one of the most common approach to calculate the unpolarized nonperturbative FFs is the phenomenological approach based on the data analysis. In order to present the theoretical predictions for  the observables involving  cross section of identified hadrons in the final state, considering different hierarchical features is vital which is mentioned in this section. In this regard,  we first review the QCD framework including  the standard factorization theorem for differential cross section in a hadronization process of single inclusive electron-positron annihilation. We shall also introduce the {\tt OPAL} and {\tt Belle} experimental data as the only data sets for $\Lambda_c^+$ production in SIA process. Finally,  we indicate our theoretical formalisms for determination of $\Lambda_c^+$ FFs and describe our strategy to determine the uncertainties of FFs as well as  corresponding theoretical cross sections.

\subsection{QCD framework for  $\Lambda_c^+$ baryon  FFs}\label{sec:QCD frame}

Our analysis depends on the normalized differential  cross section $1/\sigma _{tot}\times d\sigma/dx_{\Lambda}$ of the  annihilation process 
\begin{eqnarray}\label{process}
e^+e^- \rightarrow (\gamma ^*,Z)\rightarrow \Lambda_c^+ +X,
\end{eqnarray}
where, $X$ refers to the unobserved hadrons in the final state and  $\Lambda_c^+$ is the identified hadron. As usual, the scaling variable $x_{\Lambda}$ is defined as $x_{\Lambda}=2p_\Lambda\cdot q/q^2$ where,  $p_\Lambda$ and $q$ refer to the four-momentum of detected baryon and intermediate gauge boson, respectively, so that  $s=q^2$ is the collision energy. In the center of mass (CM) frame, the scaling variable is simplified as $x_{\Lambda}=2E_{\Lambda}/\sqrt {s}$ where $E_{\Lambda}$ shows the energy of detected baryon. 

The key implement to divide the perturbative and nonperturbative parts of the $e^+e^-$ annihilation process (\ref{process}) is  the factorization theorem in the QCD-improved parton model~\cite{Collins:1998rz}. According to this theorem, the differential cross section of process (\ref{process}) is written as the convolution of differential partonic cross sections $d\sigma_i(e^+e^-\to i+X)/dx_i$, with the $\Lambda_c^+$-FFs which is denoted by $D_i^{\Lambda_c^+}$. Here,  $i=g,u,\bar{u},\ldots,b,\bar{b}$ runs over the active partons so that the number of active flavors is dependent on the energy scale.
In the ZM-VFNS (or zero-mass scheme)  where all light and heavy quarks are considered as massless partons, the differential cross section normalized to the total one  is written as \cite{Collins:1998rz}
\begin{align}\label{differentional cross}
\frac{1}{\sigma _{tot}}&\frac{d\sigma}{dx_{\Lambda}}(e^+e^- \rightarrow \Lambda_c^+ +X)\nonumber\\
&=\sum _{i} \int ^{1}_{x_{\Lambda}} \frac {dx_i}{x_i}\frac{1}{\sigma_{tot}}\frac{d\sigma _i}{dx_i}(x_i,\mu _R, \mu _F) D^{\Lambda_c^+}_{i}(\frac{x_\Lambda}{x_i},\mu _F).
\end{align}
In the CM frame, the scaling variable $x_i$ is also defined as $x_i=2E_i/\sqrt{s}$ which refers to  the energy of produced parton $i$ in units of the beam energy.  In the relation above, the scales $\mu _F$ and $\mu _R$ are the factorization and renormalization scales, respectively. Normally, they are arbitrary quantities which appear in each order of perturbation but in order to omit the ambiguous logarithmic terms $\ln (s/\mu _F^2)$ in the partonic cross sections,  they are chosen to be $\mu _F=\mu _R=\sqrt{s}$.\\
The experimental data included in our analysis are normalized to the total hadronic cross section for the $e^+e^-$ annihilation. This cross section reads 
\begin{eqnarray}
\label{total}
\sigma _{tot}&&=\frac{4\pi \alpha^2(s)}{s}(\sum ^ {n_f}_i \tilde{e}^2_i(s))\nonumber\\
&& \times (1+\alpha _sK^{(1)}_{QCD}+\alpha _s ^2 K^{(2)}_{QCD}+\cdots).
\end{eqnarray}
Here, $\alpha_s$ and $\alpha$  are the strong-coupling and fine-structure constants, respectively, and $\tilde{e}_i$ is the effective electroweak charge of
quark $i$.
The QCD perturbative coefficients $K^{(i)}_{QCD}$ are currently known up to NNLO accuracy \cite{Gorishnii:1990vf} so that  $K_{\rm QCD}^{(1)}=3C_F/(4\pi)$, where $C_F=4/3$, and $K_{\rm QCD}^{(2)}\approx1.411$
\cite{Chetyrkin:1979bj}. \\
In Eq.~(\ref{differentional cross}),  the nonperturbative part of the process (\ref{process}) related to the transition $i\to \Lambda_c^+$ is  described by the $D_i^{\Lambda_c^+}(z, \mu_F)$-FF, where the fragmentation parameter $z = x_\Lambda/x_i$ indicates the energy fraction passed on from parton $i$ to the $\Lambda_c^+$ baryon, i.e., $z=E_\Lambda/E_i$. 
Since the FFs depend on the factorization scale $\mu _F$, they are evaluated to the various scales of energies by the DGLAP evolution equations, i.e.  
\begin{eqnarray}
\label{DGLAP}
\mu ^2 \frac{dD_i^{\Lambda}}{d\mu ^2}(x_{\Lambda},\mu)=\sum _ j\int ^1_{x_\Lambda} \frac{dx_i}{x_i}P_{ij}(\frac{x_{\Lambda}}{x_i},\alpha _s(\mu))D_j^{\Lambda}(x_i,\mu),
\end{eqnarray}
where, $P_{ij}$ are the splitting functions which have been computed up to NNLO accuracy \cite{Mitov:2006ic,Moch:2007tx,Almasy:2011eq}.

\subsection{Theoretical formalism}\label{sec:error}

According to the phenomenological approach, in order to extract  the nonperturbative FFs, the $z$-distributions of $i\rightarrow \Lambda_c^+$ FFs at the starting scale $\mu_0$ are parametrized from the beginning and the free parameters are constrained from the SIA experimental data. 
Note that, the selection criterion for the best parametrization form is to score a minimum $\chi^2_{\tt global}$ value as small as
possible with a set of fit parameters as minimal as possible. The $\chi^2_{\tt global}$ function is defined in our previous works in more details~\cite{Soleymaninia:2017xhc,Salajegheh:2019ach}.  Following Ref.~\cite{Kniehl:2020szu},
we parametrize the $z$-dependence of $c\to \Lambda_c^+$ and  $b\to \Lambda_c^+$ FFs at the starting scale as suggested by Bowler \cite{Bowler:1981sb}  while the FFs of light flavors are assumed to be zero at the starting scale. 
Moreover, since the data set included in our analysis is limited to the SIA process we can not constrain the gluon FF at the initial scale,  thus  its corresponding FF is also  set equal to zero at the initial scale, i.e., 
\begin{eqnarray}
\label{light FFs}
D^{\Lambda_c^+}_i(z,\mu _0^2)=0,~~ for~ i=u,\bar{u},d,\bar{d},s,\bar{s},g.
\end{eqnarray}
The FFs of gluon and light flavors will be generated  to higher energy scales via the DGLAP evaluation equations. 
As was mentioned, for the  $c\rightarrow \Lambda_c^+$ and $b\rightarrow \Lambda_c^+$ fragmentation the following Bowler parametrization \cite{Bowler:1981sb} is considered:
\begin{eqnarray}
\label{FF parameters}
D^{\Lambda_c^+}_c(z,\mu _0^2)=N_c(1-z)^{a_c}z^{-(1+b_c^2)}e^{-b_c ^2/z} \nonumber \\
D^{\Lambda_c^+}_b(z,\mu _0^2)=N_b(1-z)^{a_b}z^{-(1+b_b^2)}e^{-b_b ^2/z},
\end{eqnarray}
which includes  six free parameters: $N_c, a_c$, $b_c,N_b, a_a$ and $b_b$. It is found that the Bowler parametrization enables one to do excellent fits at each order of perturbation, i.e. LO, NLO and NNLO. Here,  the initial scale is set as $\mu_0 =4.301$~GeV which is a little grater than the bottom mass threshold $m_b=4.3$~GeV.  
\\
Consequently, we have six free parameters which should be extracted from the best QCD fit on the experimental data.  
The optimal values of fit parameters  for the charm and bottom FFs  are reported in Table \ref{tab1} at each order of perturbation. 
\begin{table}[h!]
	\caption{ The optimal values for the input parameters of the  $c\to \Lambda_c^+$ and  $b\to \Lambda_c^+$ FFs at the initial scale $\mu_0^2 = 18.5$~GeV$^{2}$ determined by  QCD analysis of the experimental data listed in Table~\ref{tab2}.}
	\label{tab1}
	\begin{ruledtabular}
		\begin{tabular}{lclll}
			Parameter & & Best values &   \\   & {\tt LO}  &   {\tt NLO}  &   {\tt NNLO} \\ \hline
			$N_c$						& $ 1928924455.183$ & $15210639.353$& $2720317.971$ \\
			$a_c$                                                 & $2.343$ & $2.373$& $2.458$ \\
			$b_c$			                         & $4.343$ & $3.814$& $3.597$ \\
			$N_b$						& $578.441$ & $323.094$ & $318.980$  \\
			$a_b$ 				                 & $9.302$  & $9.138$ & $9.321$ \\
			$b_b$ 					         & $1.554$  & $1.477$ & $1.464$
		\end{tabular}
	\end{ruledtabular} 
\end{table}

Technically, it should be mentioned that  for the evolution of $D^{\Lambda^+_c}_i(z, \mu^2)$-FFs as well as for the calculation of  SIA cross sections up to NNLO accuracy  we employed the publicly available {\tt APFEL} package~\cite{Bertone:2013vaa} and the free parameters of FFs are determined by minimizing the  $\chi^2_{\tt global}$ function using the CERN program {\tt MINUIT}~\cite{James:1975dr}.  
Furthermore, to estimate the  uncertainties of $D^{\Lambda^+_c}_i(z, \mu^2)$-FFs 
the  data uncertainties are propagated to the extracted QCD fit parameters using the asymmetric Hessian method (or Hessian methodology), as is outlined in~\cite{Martin:2009iq, Stump:2001gu}. More details can be found in our previous analysis \cite{Salajegheh:2019ach}.
  
\subsection{Experimental data and fit results}\label{sec:error}

In our analysis, we applied two data sets measured by {\tt OPAL} Collaboration at the CERN LEP1 Collider \cite{Alexander:1996wy}. 
In fact, for the SIA process (\ref{process}) two different mechanisms contribute with similar rates; direct production through $Z \rightarrow c\bar{c}$ decay followed by the fragmentation $c/\bar{c}\to \Lambda_c^+$ and the decay $Z \rightarrow b \bar{b}$ (b-tagged events)  followed by the fragmentation of $b$ (or $\bar b$) into the bottom-flavored hadron $H_b$, i.e., $b/\bar{b}\to H_b$, where finally the  weak decay of $H_b$ into the $\Lambda_c^+$-baryon occurs; $H_b\to \Lambda_c^++X$. Consequently, the energy spectrum of $\Lambda_c^+$-baryon  originating from the decay of $H_b$-hadron is much softer than that due to primary charm production, as is expected. In order to separate charmed hadron production through the decay process $Z \rightarrow c \bar{c}$ from the decay $Z \rightarrow b \bar{b}$, the {\tt OPAL} Collaboration investigated the apparent decay length distributions as well as the energy spectra of  charmed hadrons. As is expected, the decay lengths of $H_b$ hadrons into $\Lambda_c^+$ baryon are always longer than those from prompt production.
Note that, the {\tt OPAL} Collaboration has presented $x_\Lambda$-distributions for their $\Lambda_c^+$ sample and for the b-tagged ($Z\to b\bar{b}$) subsamples.  In addition to the {\tt OPAL} data which includes only 4 points with rather large uncertainties we have also included a very recent data set measured by {\tt Belle} Collaboration  \cite{Niiyama:2017wpp} at $\sqrt{s}=10.52$~GeV. This new data set is much more precise and contains more points. 
{\tt Belle} data dose not have contributions from B-meson decays so that the contribution of $b\to \Lambda_c$ is not needed to be taken in the calculation of cross sections. However, the FFs from charm and bottom quarks are coupled through the DGLAP evolution equation.
Following Ref.~\cite{Kniehl:2020szu}, we fix the $b\to \Lambda_c$  FF using the values of $N_b, a_b$ and $b_b$ extracted from the {\tt OPAL} fit. Therefore, the values of $N_c, a_c$ and $b_c$ are yielded from the fit to the {\tt Belle} data. 
Here, we describe a technical point over the reconstruction  of  old {\tt OPAL} experimental data. {\tt OPAL} data sets have been displayed in the form $(1/N_{\tt had}) dN/dx_\Lambda$  where $N$ refers to the number of charmed-flavor hadron candidates which are reconstructed through appropriate decay chains.  Therefore, in order to convert these data into the convenient cross section  $(1/\sigma_{\tt tot})d\sigma/dx_\Lambda$, it is needed to  divide them by the corresponding branching fractions of decays for the reconstruction of  charmed-flavored baryons. In Refs.~\cite{Alexander:1996wy,Ackerstaff:1997ki},  for the required branching fraction the following value is applied:
\begin{align}\label{eq5}
Br (\Lambda_c^+ \rightarrow
pK^-\pi^+) =  
(4.4 \pm 0.6)\%.
\end{align}
Since, in our analysis  we are including both the old {\tt OPAL} data and more recent  {\tt Belle} data which  are based on the observation of decay $\Lambda_c^+ \rightarrow pK^-\pi^+$, then we have to use the same branching ratio to reconstruct both data sets. Since, the {\tt Belle} analysis has applied the newest branching ratio as $Br (\Lambda_c^+ \rightarrow pK^-\pi^+) =  (6.635)\%$  \cite{Patrignani:2016xqp}, we therefore rescale  the old {\tt OPAL} data    by the factor $0.044/0.0635=0.6929$.

\begin{figure*}[t]
	\includegraphics[width=0.5\linewidth, angle =-90]{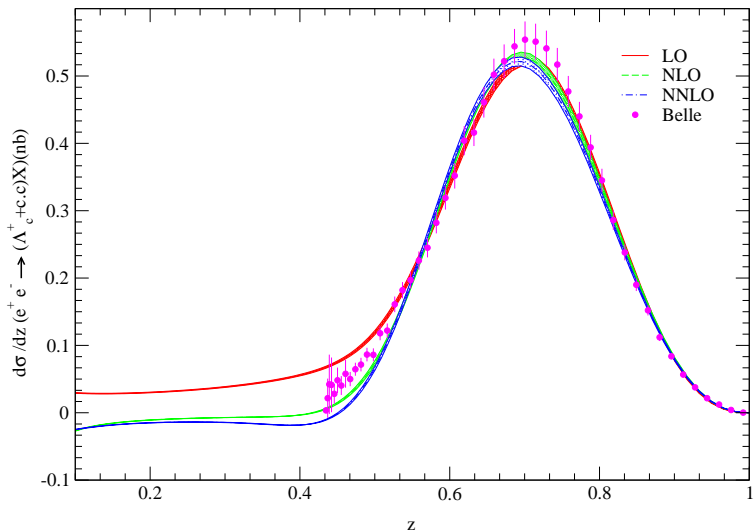}
		\begin{center}
		\caption{{\small {Using our extracted $\Lambda _c^+$-FFs, the theoretical predictions for inclusive differential cross sections at LO (red solid line), NLO (green dashed line) and NNLO (blue dot-dashed line) in $\sqrt{s}=10.52$ GeV are compared with {\tt Belle} experimental data points fitted to. Corresponding uncertainty bands stem from $\Lambda _c^+$ FFs are shown as well. } \label{fig1}}}
	\end{center}
\end{figure*}

\begin{figure*}[t]
        \vspace{-1cm}
        	\includegraphics[width=0.5\linewidth, angle =-90]{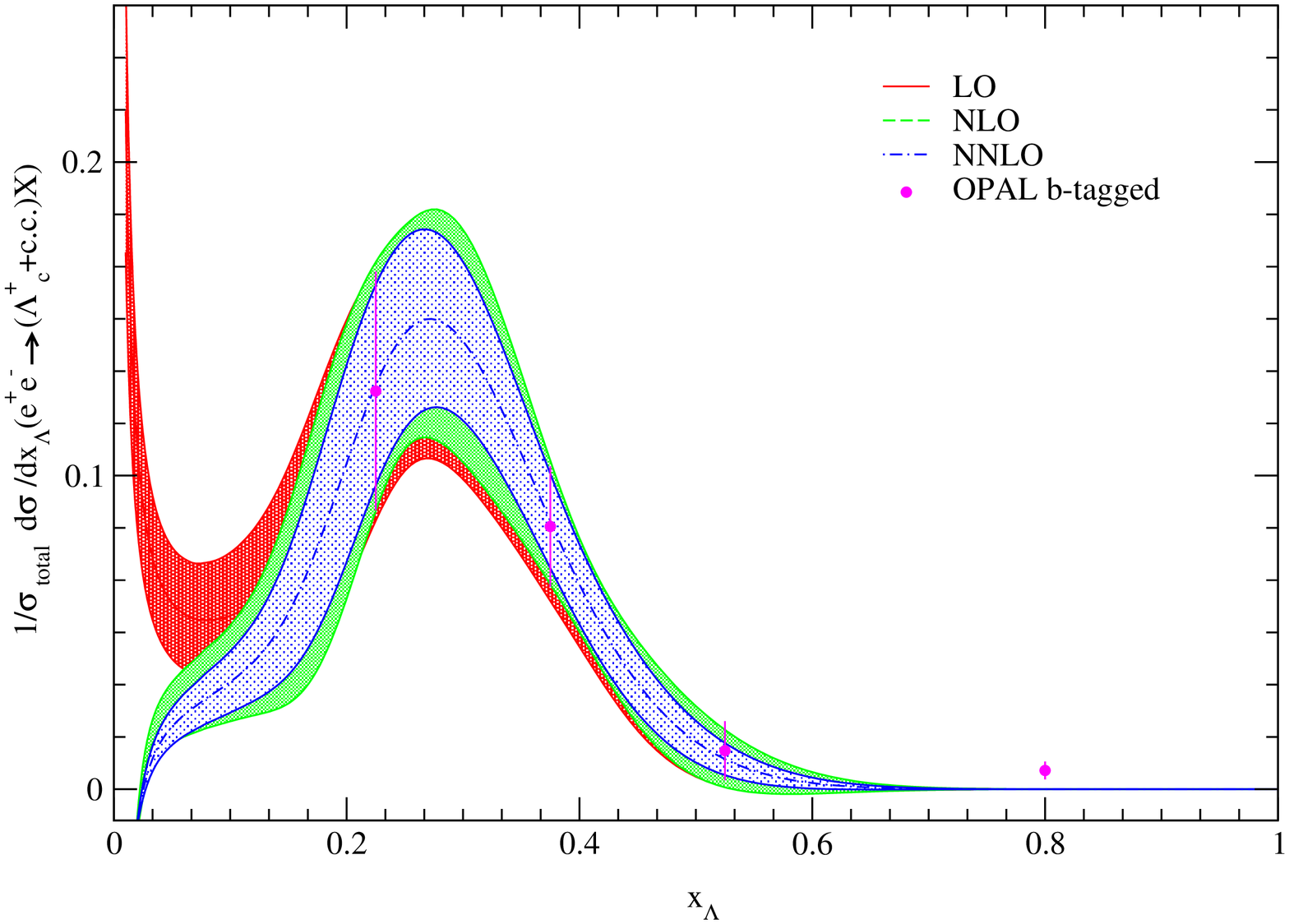}
	
	\vspace{-1cm}
	\includegraphics[width=0.5\linewidth, angle =-90]{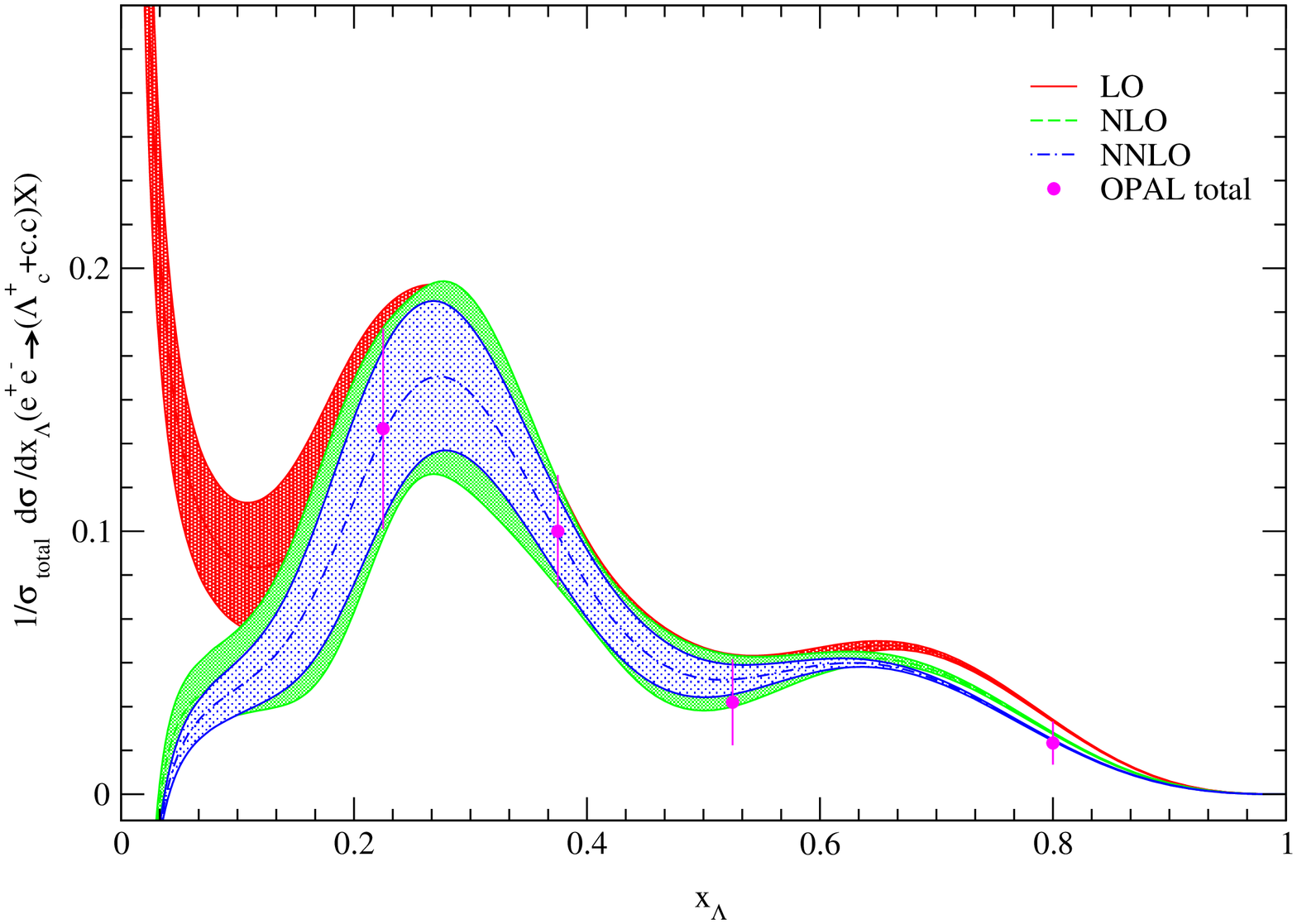}
	
	\begin{center}
		\caption{{\small {Same as Fig.~\ref{fig1} but for {\tt OPAL} experimental data at $\sqrt{s}=M_Z$. } 
		\label{fig2}}}
	\end{center}
\end{figure*}

\begin{figure*}[t]
	
	\includegraphics[width=0.6\linewidth, angle =-90]{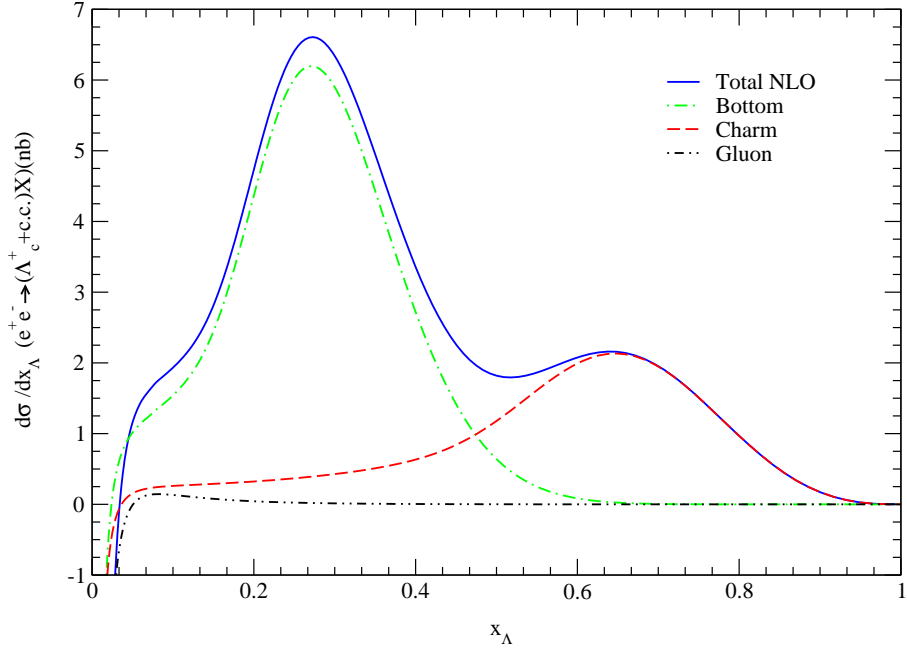}
	\vspace{-1cm}
	\begin{center}
		\caption{{\small {The NLO contribution of $ c\rightarrow \Lambda _c ^+$ (red dashed line), $ b\rightarrow \Lambda _c ^+$ (green dot-dashed line) and $g \rightarrow \Lambda _c ^+$ (black dashed-dot-dot line)  in inclusive differential cross section at $\sqrt{s}=M_Z$. The total contribution of all partons (blue solid line) is also plotted.
				 } \label{fig3}}}
	\end{center}
\end{figure*}

\begin{figure*}[t]
	\includegraphics[width=0.8\linewidth, angle =-90]{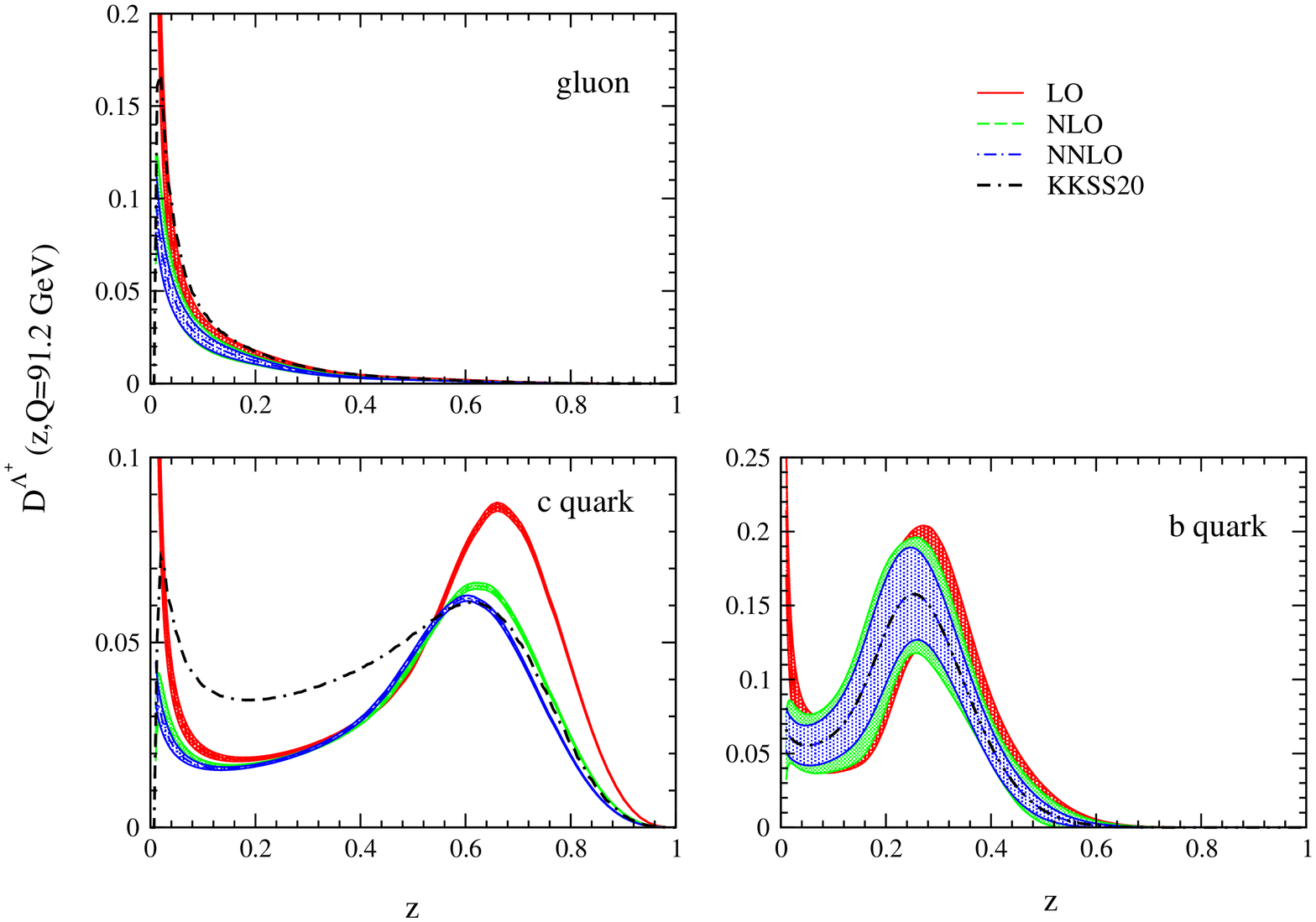}
	\vspace{-1cm}
	\begin{center}
		\caption{{\small {The gluon-, charm- and bottom-FFs with their uncertainties obtained  at LO (red solid lines), NLO (green dashed lines) and NNLO (blue dot-dashed lines) QCD analyses of $\Lambda_c^+$ baryon productions at $\mu =M_Z$. The NLO {\tt KKSS20} results \cite{Kniehl:2020szu} (black dashed-dashed-dot curves) are also plotted. } \label{fig4}}}
	\end{center}
\end{figure*}

Another point about the {\tt Belle} data is that we have to exclude data at small values of the scaling variable $z\le 0.5$, since the theory is not reliable in this range without taking resummation of soft-gluon logarithms into account. Note that, no kinematical cut is taken over  the {\tt OPAL} data.
\begin{table*}
\begin{center}
\begin{tabular}{lp{2.2cm}p{2.2cm}p{2.2cm}p{2.2cm}p{2.2cm}p{2.2cm}p{2.2cm}}
				\hline\hline
				Dataset & Observable & $\sqrt{s}$[GeV]& $N_n^{data}$ & $\chi^2$ ({\tt LO}) & $\chi^2$ ({\tt NLO}) & $\chi^2$ ({\tt NNLO})                  \\ \hline\hline
		{\tt Belle}   & Inclusive   &10.52& 35 &41.419 &42.843& 55.413  \\
		{\tt OPAL}   & Inclusive   &91.2& 4 & 1.971&0.444& 0.299  \\
		& $b$-tagged  &91.2&  4 & 4.520 & 4.524& 4.539 \\ \hline
		{\bf TOTAL:}  &&& 43 &  47.948  &47.812&   60.251 \\
		($\chi^{2}$/{ d.o.f})  &&& &1.296 &1.292 & 1.628        \\\hline\hline

\end{tabular} 		
\end{center}
\caption{ The individual $\chi ^2$ values for inclusive and $b$-tagged cross sections
		obtained at LO, NLO and NNLO. The total $\chi ^2$ and $\chi ^2/d.o.f$ fit for $\Lambda_c^+$ are also shown. }
\label{tab2}	
\end{table*}
In Table.~\ref{tab2}, the characteristics of available experimental data including the number of data points $N_n^{data}$ are presented. 
We have also listed the individual values of $\chi^2$ for inclusive and $b$-tagged data sets at LO, NLO and NNLO accuracies. 
Considering the number of degrees of freedom (d.o.f), i.e., $43-6=37$, 
we have also presented the total $\chi^2$ divided
by the number of d.o.f at LO ($\chi^2/d.o.f =1.296$), NLO ($\chi^2/d.o.f =1.292$) and NNLO ($\chi^2/d.o.f =1.628$). 
As is seen, these values  are around 1 for individual data sets so this confirms a well-satisfying fit in all three accuracies.
From Table.~\ref{tab2}, it is seen that a reduction in $\chi^2/d.o.f$  occurs when passing from LO to NLO accuracy. This behavior is not valid when NNLO corrections are considered.  
This is due to the fact that, the {\tt Belle} cross sections have been measured at small scale of energy ($\sqrt {s}=10.52$ GeV) where the corrections for  the finite mass of hadron and partons  are more effective than the higher order corrections. Remember that we have used the ZM-VFN scheme where the hadron and parton masses are set to zero from the beginning. Nevertheless, the NNLO corrections lead to a reduction in the uncertainties bands  of FFs and corresponding observables. See Figs.~\ref{fig1}-\ref{fig4}.   

In order to show the consistency and goodness between the theoretical prediction and the experimental data used in the fits, in Fig.~\ref{fig1} we have  plotted  the inclusive differential cross section, as reported by {\tt Belle}, and  in Fig.~\ref{fig2} we have plotted the b-tagged and total differential cross sections  normalized to the  total  one evaluated with our respective $\Lambda_c^+$ FFs. Both are compared with the {\tt Belle} and {\tt OPAL}  data sets fitted to. 
In these figures, the uncertainty bands are also plotted using the  Hessian approach. Through this approach, we just  considered the uncertainties due to  the experimental data sets so that  we ignored additional sources of uncertainties.  
As is seen the quality of fit is improved when passing to higher order corrections. 
From Fig.~\ref{fig2}, it is seen that our theoretical descriptions at LO, NLO and NNLO for both b-tagged and total differential cross sections are in good mutual agreement.  The consistency seems to be better for the normalized total differential cross section (shown in lower panel) in comparison with the b-tagged one because our theoretical predictions do not go across one of the b-tagged data points  located at $x_\Lambda=0.8$. This is why  higher values of individual $\chi ^2$ occur for b-tagged, see Table.~\ref{tab2}. 

In Fig.~\ref{fig2}, it is also seen that  the behavior of theoretical predictions in small values of $x_\Lambda$ for the lowest order accuracy  is completely different with the ones at NLO and NNLO. Obviously, in one hand, the  theoretical cross section at LO  goes to infinity when $x_\Lambda\to 0$ and, on the other hand, the LO uncertainty band  is anomalously wider than the NLO and NNLO ones in all ranges of $x_\Lambda$. Therefore, the LO results are not reliable so that higher order radiative corrections need to be considered.  
\\
In order to show the fragmentation contribution of gluon, charm- and bottom-quark to the production of $\Lambda _c^+$, in Fig.~\ref{fig3} we have plotted these contributions at the scale $\sqrt{s}=M_Z$.  The total  differential cross section at NLO  is also shown which is obtained by the sum of all contributions. As is expected, the  contribution of gluon fragmentation is very tiny and it increases at small range of $x_\Lambda$.  At large $x_\Lambda$, the contribution of charm quark (red dashed line) is governed while at small region the contribution of bottom quark (green dot-dashed line) is governed.

In Fig.~\ref{fig4}, the $z$-distributions of  $\Lambda_c^+$-FFs  are plotted at $\mu _F=M_Z$; the energy scale of {\tt OPAL} data sets fitted to. 
For this purpose, we plotted  the $(c, b, g)\to D_{\Lambda^+_c}$ FFs at LO (solid lines), NLO (dashed lines) and NNLO (dot-dashed lines).
From this plot, it is seen that the fragmentation of charm-quark is peaked at large-$z$  whereas the bottom fragmentation  has its maximum at small-$z$. This behavior is due to the fact that the fragmentation process $b\to \Lambda_c^+$ contains  two-step mechanism.
In Fig.~\ref{fig4}, the uncertainty bands of $\Lambda_c^+$-FFs are also presented which are needed to visually quantify the remaining error of analysis. Since, the {\tt Belle} date
does not include the contributions from the $b\to \Lambda_c$ fragmentation  in the calculation of cross sections, then the uncertainties of $b$-quark FF are considerably much wider  than the charm and gluon ones in each order of perturbation. Moreover,  the error bands of all flavors decrease by increasing the order of perturbation. 
In Fig.~\ref{fig4}, the NLO {\tt KKSS20} results \cite{Kniehl:2020szu} (dashed-dashed-dot curves) are also plotted. 
As is seen, our result for gluon fragmentation is in good agreement with the one presented by {\tt KKSS20}. In comparison to the {\tt KKSS20}'s results, there is a considerable difference between the {\tt KKSS20} charm FF and ours in the range $z<0.5$. However, the behavior of our bottom-FF  is the same as the {\tt KKSS20} one in the whole range of $z$. Unlike our procedure in which we set the same scale for the c- and b-quark FFs, the {\tt KKSS20} collaboration has selected different initial scales so that in their work 
the starting scales for the charm- and bottom-FFs were taken to be $\mu_0=m_c=1.5$~GeV and $\mu_0=m_b=5$~GeV, respectively.

\section{Theoretical approach for $\Lambda_c^+$ baryon FF} \label{sec:theoretical}

As was mentioned in the Introduction, apart from the phenomenological approaches to determine the nonperturbative FFs there are also some theoretical models to compute them. In fact, it was fortunately understood that for heavy hadron productions these functions can be analytically calculated by virtue of perturbative QCD (pQCD) including limited phenomenological 
parameters \cite{Ma:1997yq,Braaten:1993mp}. The first theoretical effort to illustrate the procedure of heavy hadron production was established by Bjorken \cite{Bjorken:1977md}, so that in the following,  Suzuki \cite{Suzuki:1977km},  Ji and Amiri \cite{Amiri:1986zv} have applied the pQCD approach considering elaborate models to describe the hadronization process. Since the Suzuki model includes most of the kinematical and dynamical aspects of hadroproduction  process it gives us a detailed insight  on the hadronization  process. Especially, this model is much suitable to consider the spin effects of produced hadron or fragmenting parton  which is absent in the phenomenological approach, see for example \cite{Nejad:2015far}. It does also enable us to describe the gluon hadronization process which is not well determined in the phenomenological approach, see Refs.~\cite{Nejad:2015oca,Delpasand:2020mqv,Delpasand:2019xpk}. \\
\begin{figure}[t]
\begin{center}
\subfigure[]{
\centering
\includegraphics[width=0.45\linewidth]{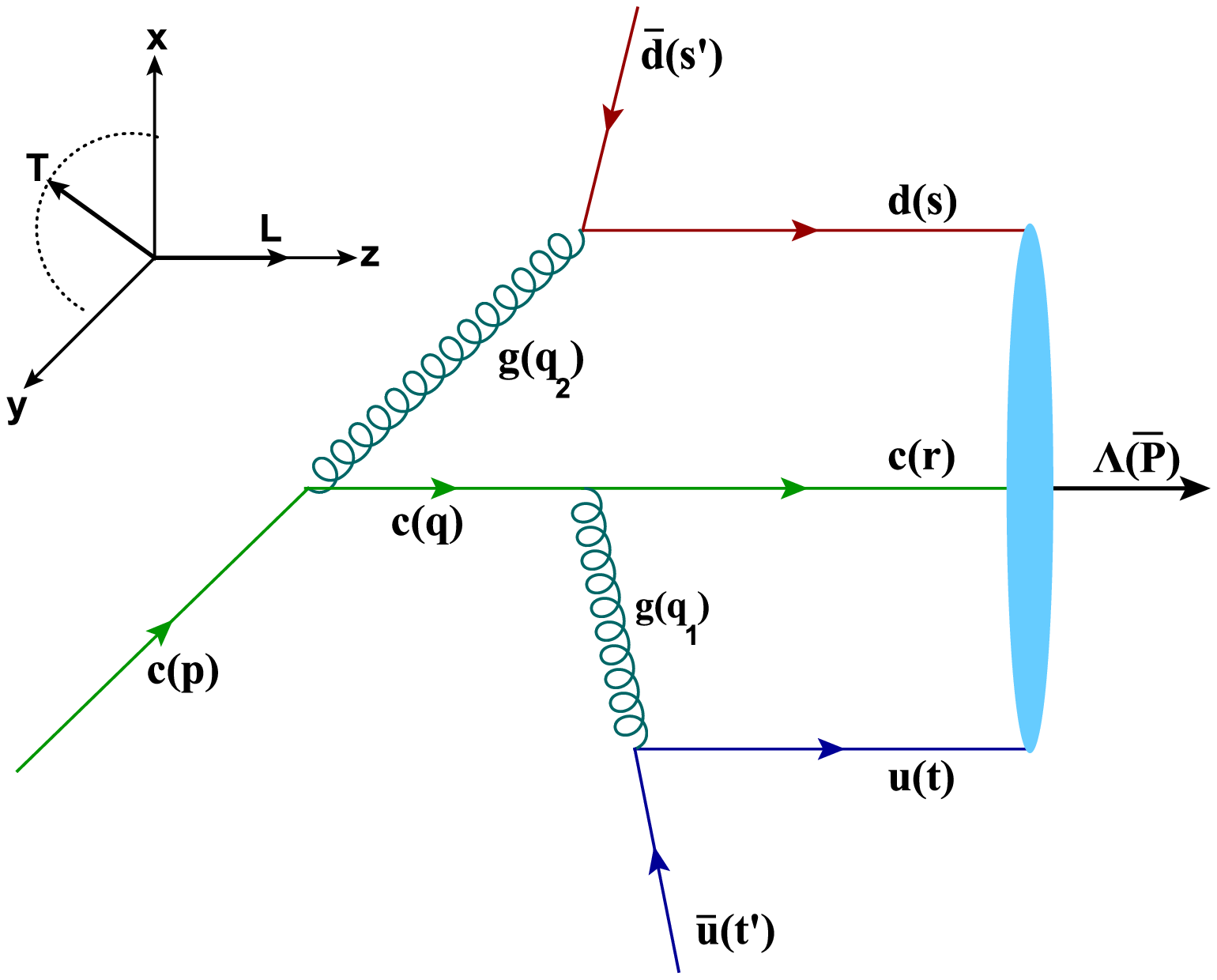}
\label{a}
}
\subfigure[]{
\centering
\includegraphics[width=0.45\linewidth]{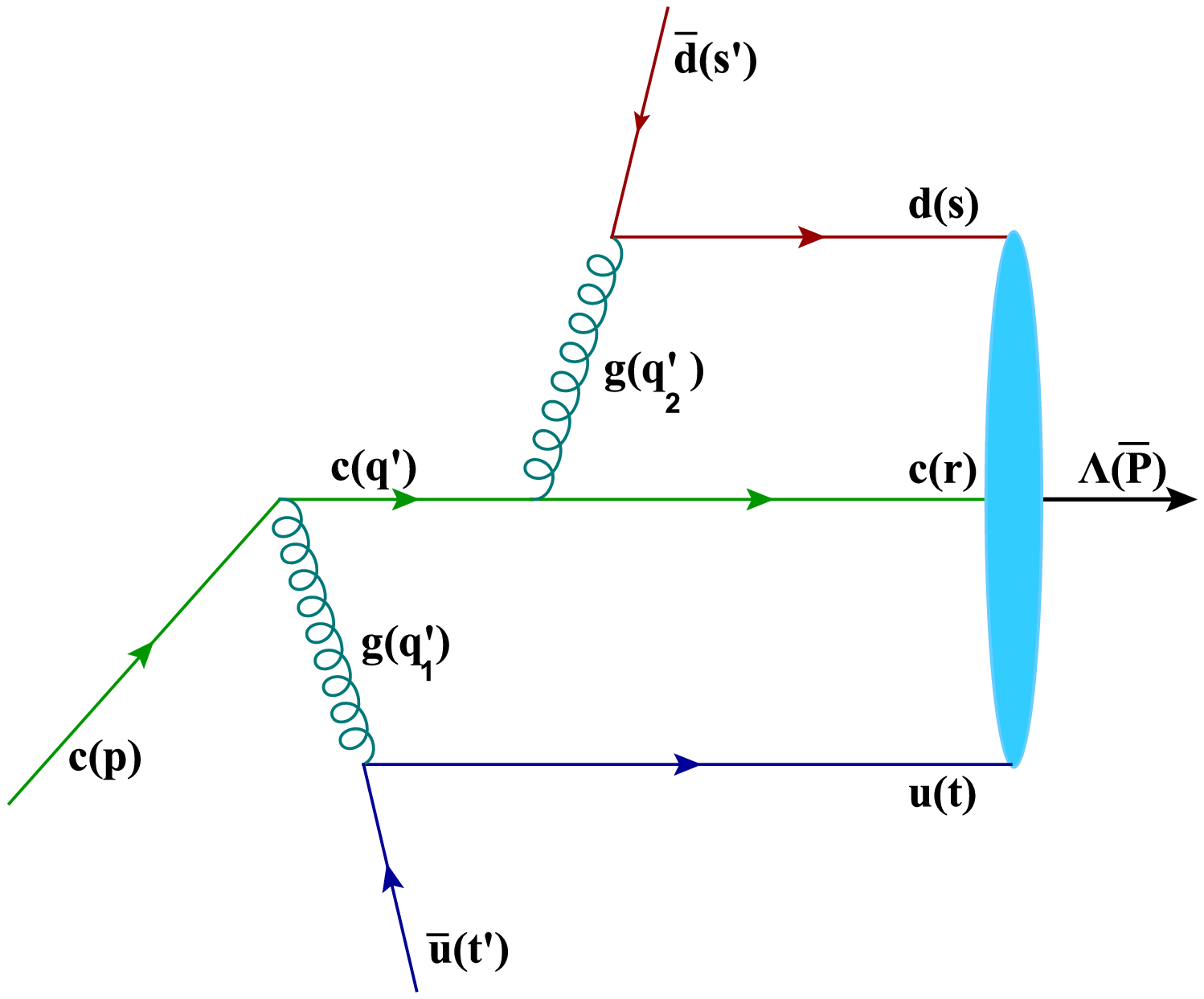}
\label{b}
}
\caption{\label{pQCD}%
			The lowest-order Feynman diagrams contributing to the fragmentation of charm-quark into the  $\Lambda_c^+(ucd)$ baryon.
}
\end{center}
\end{figure}
In this section, using the Suzuki model we focus on the fragmentation of  $\Lambda_c^+$-baryon from charm quark for which the respective Feynman diagrams at leading order in $\alpha_s$ are shown in Fig.~\ref{pQCD}.
According to this model, the FF for the production of an S-wave 
heavy bound state $M$ in  hadronization of an initial heavy  quark $Q$ might  be put in the following general relation \cite{Suzuki:1977km}
\begin{eqnarray}\label{first}
D_Q^M(z, \mu_0)&=&\nonumber\\
&&\hspace{-2.2cm}\frac{1}{1+2s_Q}\sum_{\begin{array}{c} spin\\color \end{array} }\int |T_M|^2 \delta^3(\sum_f\boldsymbol{p}_f-\boldsymbol{p}_Q)\Pi_{f}d^3\boldsymbol{p}_f,  
\end{eqnarray}
where, $\boldsymbol{p}_Q$ and $\boldsymbol{p}_f$ are the momenta of the fragmenting quark and the final particles, respectively. The fragmentation  parameter $z$ is  as the one introduced in the phenomenological approach (section~II.A), i.e., $z=E_M/E_Q$ which takes the values as $0\le z \le 1$. Furthermore, $\mu_0$ stands for the initial fragmentation scale which is in order of fragmenting heavy quark mass and $s_Q$ refers to the fragmenting quark spin.  
In the above relation, the quantity $T_M$ is
the  probability amplitude for the hadron production. In the Suzuki model, this amplitude at large momentum transfer
is expressed in terms of the  hard scattering 
amplitude $T_H$  and the process-independent  distribution amplitude $\Phi_M$ describing the nonperturbative dynamics of bound state. In fact, the long-distance amplitude $\Phi_M$ is, in essence, the probability amplitude for constituent quarks to be evolved
into the final bound state. Therefore, the  amplitude $T_M$ is expressed as \cite{Amiri:1986zv,Brodsky:1985cr}
\begin{eqnarray}\label{base}
T_M=\int \Pi_j dx_j\delta(1-\sum_j x_j) T_H \Phi_M(x_j, Q^2),
\end{eqnarray}
where, $x_j=E_j/E_M$  is the energy fraction carried by the constituent quark $j$ of heavy bound state $M$.  \\
Considering the general definition (\ref{first}) and the Feynman diagrams shown in Fig.~\ref{pQCD}, for the production of S-wave  $\Lambda_c^+$-baryon  from the initial charm quark, the  FF of $c\to \Lambda_c^+$ is written as 
\begin{align}\label{suzuki FF}
D_c^{\Lambda_c^+}(z,\mu_0)&=\frac{1}{2} \sum_{s,c} \int  |T_\Lambda|^2\nonumber\\
&\times  \delta^3(\bar{\boldsymbol{P}}+\boldsymbol{s}'+\boldsymbol{t}'-\boldsymbol{p}) d^3 \bar{ \boldsymbol{P}} d^3 \boldsymbol{s}' d^3 \boldsymbol{t}',
\end{align}
where,  four-momenta are as labeled in Fig.~\ref{pQCD}, and
\begin{align}\label{probability amplitude}
T_\Lambda&(p,\bar{P},s',t')\nonumber\\
&=\int_0^1 dx_1 dx_2 dx_3\delta(1-x_1-x_2-x_3)\nonumber\\
& \hspace{1cm}\times T_H(p,\bar{P},s',t',x_i)\Phi_B(x_i,Q^2).
\end{align}
The advantage of above scheme is  to absorb the soft behavior of produced bound state into the hard scattering amplitude $T_H$  \cite{Brodsky:1985cr}. 
Ignoring the details, the distribution amplitude $\Phi_B$ is related to the valence wave function $\Psi$ \cite{Brodsky:1985cr}. Following Ref.~\cite{Suzuki:1977km}, we adopt the infinite momentum frame where
the distribution amplitude $\Phi_B$, with neglecting the Fermi 
motion of constituents, reads \cite{MoosaviNejad:2016scq,MoosaviNejad:2018ukp}
\begin{eqnarray}\label{delta}
\Phi_B= f_B \delta(x_i-\frac{m_i}{M}),
\end{eqnarray}
where, the baryon decay constant  $f_B$ is related to the nonrelativistic S-wave function $\Psi(0)$ at the origin as  $f_B=\sqrt{12/M}|\Psi(0)|$.

In Eq.~(\ref{probability amplitude}), the short-distance amplitude $T_H$  can be calculated perturbatively considering the Feynman diagrams shown in Fig.~\ref{pQCD}, 
where a charm-quark creates a heavy baryon $\Lambda_c^+$ along with two light antiquarks $\bar {u}$ and $\bar{d}$.  
In the old-fashioned noncovariant perturbation theory, the hard scattering amplitude $T_H$ may be expressed as follows to keep the initial heavy quark on mass shell all the time \cite{Suzuki:1985up} 
\begin{eqnarray}\label{hard amplitude}
T_H=\frac{g_s^4 m_\Lambda m_c m_d m_u C_F}{2\sqrt{2p_0\bar{P}_0s'_0t'_0}} \times \frac{\sum_{i=1}^2\Gamma_i}{D_0},
\end{eqnarray}
where, $D_0=\bar{P}_0+s'_0+t'_0-p_0$ is the energy denominator, $C_F$ is the usual color factor and $\Gamma_i$ represents an appropriate combination of the propagators and the spinorial parts of the amplitude. 
In the above equation, the amplitudes $\Gamma_i$ stand for each Feynman diagrams in Fig.~\ref{pQCD}.
\\
Substituting all expressions in Eq.~(\ref{suzuki FF}) and carrying out the necessary integrations, we find
\begin{align}\label{hehe}
D_c^{\Lambda_c^+}(z, \mu_0)&=N\int\frac{d^3 \boldsymbol{s}' d^3 \boldsymbol{t}'}{t_0^\prime s_0^\prime}
\int\frac{\frac{1}{2}\sum_{i, j=1}^2\Gamma_i\bar\Gamma_j}{\bar{P}_0 p_0 D_0^2}\nonumber\\
& \times\delta^3(\bar{\boldsymbol{P}}+\boldsymbol{s}'+\boldsymbol{t}'-\boldsymbol{p})d^3 \bar{ \boldsymbol{P}},
\end{align}
where, $N\propto (f_B m_\Lambda m_c m_d m_u g_s^4 C_F)^2$.\\
In the above relation, using the well-known completeness relations $\sum_{spin}u(p)\bar{u}(p)=(\displaystyle{\not}{p}+m)$ and $\sum_{spin }v(q)\bar{v}(q)=(\displaystyle{\not}{q}-m)$  in the unpolarized Dirac string, one has
\begin{eqnarray}\label{string}
\sum_{i, j=1}^2\Gamma_i\bar\Gamma_j&=&\nonumber\\
&&\hspace{-2.2cm}G_1^2Tr[(\not{s}'- m_d)\gamma^\mu(\displaystyle{\not}{s} + m_d) \gamma^\nu T_{\mu\sigma\rho\nu} (\not{t'}-m_u)\gamma^\sigma (\not{t}+m_u)\gamma^\rho]+\nonumber\\
&&\hspace{-2.2cm}G_2^2Tr[(\not{t'}-m_u)\gamma^\sigma (\not{t}+m_u)\gamma^\rho T_{\sigma\mu\nu\rho}  (\not{s'}- m_d)\gamma^\mu({\not}{s} + m_d) \gamma^\nu]\nonumber\\
&&\hspace{-2.2cm}+2G_1G_2\times\nonumber\\
&&\hspace{-2.2cm}Tr[(\not{s'}- m_d)\gamma^\mu({\not}{s} + m_d) \gamma^\nu T_{\mu\sigma\nu\rho}(\not{t'}-m_u)\gamma^\sigma (\not{t}+m_u)\gamma^\rho],
\end{eqnarray}
where,
\begin{eqnarray}
T_{\mu\sigma\rho\nu}&=& (\displaystyle{\not}{p}+m_c)\gamma_\mu (\displaystyle{\not}{q} +m_c)\gamma_\sigma (\displaystyle{\not}{r}+m_c)\gamma_\rho (\displaystyle{\not}{q} +m_c)\gamma_\nu,\nonumber\\
T_{\sigma\mu\nu\rho}&=&(\displaystyle{\not}{p}+m_c)\gamma_\sigma (\not{q'} +m_c)\gamma_\mu (\displaystyle{\not}{r}+m_c)\gamma_\nu (\not{q'} +m_c)\gamma_\rho, \nonumber\\
T_{\mu\sigma\nu\rho}&=& (\displaystyle{\not}{p}+m_c)\gamma_\mu (\not{q'} +m_c) \gamma_\sigma (\displaystyle{\not}{r}+m_c)\gamma_\nu (\not{q} +m_c)\gamma_\rho .\nonumber\\
\end{eqnarray}
In the above expressions, after using the Dirac algebra and the traditional trace technique  the dot products of four-momenta will appear. To proceed we need to specify our kinematics to determine the relevant dot products. 
Considering the Feynman diagrams shown in Fig.~\ref{pQCD}, where by ignoring the Fermi motion of quark constituents the  $\Lambda_c^+$ baryon is replaced by its collinear constituents, the relevant four-momenta are set as
\begin{eqnarray}\label{kinematic}
&&p_\mu =[p_0, \boldsymbol{p}_T, p_L],\quad    t_\mu^\prime =[t_0^\prime, \boldsymbol{t'}_T, t_L^\prime],\nonumber\\
&& s_\mu^\prime =[s_0^\prime, \boldsymbol{s'}_T, s_L^\prime], \quad r_\mu =[r_0, \boldsymbol{0}, r_L],  \nonumber\\
&& t_\mu=[t_0, \boldsymbol{0}, t_L] ,\qquad s_\mu=[s_0, \boldsymbol{0}, s_L],\nonumber\\
&&\bar P_\mu=[\bar P_0, \boldsymbol{0}, \bar P_L],
\end{eqnarray}
where, $\bar P_L=r_L+t_L+s_L$ and we also assumed that the produced baryon moves along the $\hat{z} $-axes (fragmentation axes). According to the definition of fragmentation parameter, i.e., $z=\bar P_0/p_0$, the baryon takes a fraction $z$ of the energy of initial heavy quark (each constituent a fraction of $x_1$, $x_2$ and $x_3$) and two antiquarks take the remaining $1-z$ (each one with a fraction of $x_4$ and $x_5$). Thus,  the parton energies can be expressed in terms of the initial heavy quark energy $p_0$, as
\begin{eqnarray}\label{kinematiccccs}
&&\bar{P}_0=zp_0,\quad s_0=x_1 zp_0,\quad r_0=x_2 zp_0,\quad t_0=x_3 z p_0,\nonumber\\
&&  s_0'=x_4(1-z)p_0, \quad t_0'=x_5(1-z)p_0,
\end{eqnarray}
where, the condition $x_1+x_2+x_3=1$ holds as well as $x_4+x_5=1$. Moreover, according to our assumption that baryon moves along the $\hat{z} $-axes, the transverse momentum of initial quark is carrying by two antiquarks so that in the infinite momentum frame we have $s'_T=x_4 p_T$ and $t'_T=x_5 p_T$. With the approximation (\ref{delta}), we are postulating that 
the contribution of each constituent  from the baryon energy  is proportional to its mass, namely, $x_i=m_i/M$ where $M=m_u+m_d+m_c$ . We also assume that $x_4=m_d/m'$ and $x_5=m_u/m'$ where $m'=m_d+m_u$.

Regarding the  kinematics introduced,  the dot products of relevant four-momenta  read
\begin{eqnarray}\label{mohsen}
s'\cdot t'&=&s\cdot t=m_u m_d,\nonumber\\
s\cdot r&=& m_c m_d\quad ,\quad t\cdot r= m_c m_u,\nonumber\\
p\cdot s'&=& \frac{m_d}{2}\beta \quad ,\quad  p\cdot t'=\frac{m_u}{2}\beta,\nonumber\\
p\cdot s&=& \frac{m_d }{2}\eta \quad ,\quad p\cdot r =\frac{m_c }{2}\eta,\nonumber\\
p\cdot t&=& \frac{m_u }{2}\eta \quad ,\quad s\cdot s'=\frac{m_d^2 }{2}\alpha,\nonumber\\
t\cdot t'&=& \frac{m_u^2}{2}\alpha \quad ,\quad r\cdot t'= \frac{m_u m_c}{2}\alpha,\nonumber\\
r\cdot s'&=& \frac{m_d m_c}{2}\alpha \quad ,\quad  s\cdot t'=\frac{m_d m_u}{2}\alpha 
\end{eqnarray}
where,
\begin{eqnarray}  
\eta&=&\frac{M}{z}+\frac{zm_c^2}{M}(1+\frac{p_T^2}{m_c^2}),\\
\alpha&=&\frac{zm'}{M(1-z)}(1+\frac{p_T^2}{m'^2})+\frac{(1-z)M}{zm'},\nonumber\\
\beta&=&\frac{m_c^2(1-z)}{m'}(1+\frac{p_T^2}{m_c^2})+\frac{m'}{1-z}(1+\frac{p_T^2}{m'^2})-2\frac{p_T^2}{m'}.\nonumber
\end{eqnarray}
In Eq.~(\ref{string}), $G_1$ and $G_2$ are related to the denominator of propagators as 
\begin{eqnarray}  
G_1&=&\frac{1}{m_d^3 m_u^2 (2 + \alpha)^2(m_d (2+\alpha)-\eta-\beta)},\nonumber\\
G_2&=&\frac{1}{m_d^2 m_u^3 (2 + \alpha)^2(m_u (2+\alpha)-\eta-\beta)}.
\end{eqnarray}
For the phase space integrations in the relation (\ref{hehe}), one has
\begin{eqnarray}\label{ayda1}
&&\int \frac{d^3\bar{\boldsymbol{P}}\delta^3(\bar{\boldsymbol{P}}+
	\boldsymbol{t^\prime}+\boldsymbol{s'}-\boldsymbol{p})}{\bar{P}_0 p_0 D_0^2}=\frac{z}{G^2(z)},
\end{eqnarray}
where,  $G(z)=M^2-m_c^2-m_u^2-m_d^2-2t^\prime\cdot s^\prime+2p\cdot t^\prime+2p\cdot s^\prime=m'(2m_c+\beta)$, so that for the remaining integrals, according to the Suzuki model and for simplicity, we replace the transverse momentum integrations by their average values, e.g.,
\begin{eqnarray}\label{ayda2}
\int d^3 \boldsymbol{t^\prime}\frac{f(z,t_T^{\prime 2})}{t_0^\prime}\approx m_u^2 f(z, \left\langle t_T^{\prime 2}\right\rangle),
\end{eqnarray}
where, according our assumption one has $t_T^{\prime}=x_5p_T=(m_u/m')p_T$. \\
Substituting all in Eq.~(\ref{hehe}),  the hadronization process $c\to \Lambda_c^+$ is described by the following function
\begin{align}\label{final FF}
D_c^{\Lambda_c^+}(z,\mu_0)=&N\frac{z \times \sum_{i, j=1}^2\Gamma_i\bar\Gamma_j}{ \left[2m_c+\beta(z) \right]^2}\Bigg|_{p_T^2\to \left\langle p_T^2\right\rangle},
\end{align}
where,  $N\propto (f_B m_\Lambda m_c m_d^2 m_u^2 g_s^4 C_F)^2/m'^2$ but it is determined via
$\int_0^1 D_c^{\Lambda_c^+}(z, \mu_0) dz=1$  (normalization condition) \cite{Amiri:1986zv}.
\\
In the above relation, $\sum_{i, j=1}^2\Gamma_i\bar\Gamma_j$ is given in (\ref{string}) which is simplified in terms of dot products of four-momenta after using the Dirac algebra. Due to the lengthy and cumbersome expression for this term  we ignore to present analytical result and just show our numerical analysis.  Note that, in the Suzuki model  the fragmentation function depends on both the  fragmentation parameter $z$ and the phenomenological parameter $\left\langle p_T^2\right\rangle$. Although, the $z$-dependence of FFs  is not yet calculable at each desired scale, but once they are computed at the  initial fragmentation scale $\mu_0$, their $\mu$ evolution is determined through the DGLAP  equations ~\cite{DGLAP}.
In the Suzuki model, the initial scale is the minimum value of the invariant mass of the fragmenting parton. Therefore, the FF presented in Eq.~(\ref{final FF}) should be regarded as a model for the $c\to \Lambda_c^+$ transition  at the initial scale  $\mu_0=m_\Lambda+m_u+m_d$. 

For our numerical analysis, we adopt the  input parameters as $m_c=1.43~\textrm{GeV}, m_d=4.67~\textrm{MeV}, m_u=2.16~\textrm{MeV}, f_B=0.25~\textrm{GeV}$, and $\alpha_s(m_c)=0.38$ \cite{Tanabashi:2018oca}.
The color factor $C_F$ is calculated using the simple color line counting rule so we applied $C_F = 7/6$ for our purpose.

\begin{figure}[t]
	\begin{center}
		\includegraphics[width=1\linewidth]{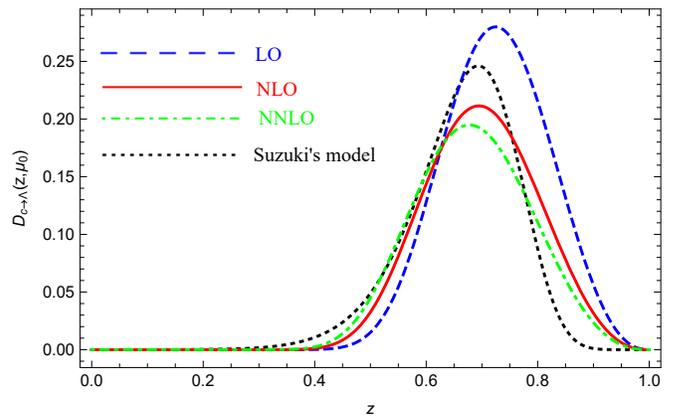}
		\caption{\label{compare}%
			The $D_c^{\Lambda_c^+}(z, \mu_0)$-FF at lowest-order in the Suzuki model (black dotted line)  taking $\left\langle p_T^2\right\rangle=1$~GeV$^2$. This is compared with the ones extracted through the phenomenological approach at LO (blue dashed line), NLO (red solid line) and NNLO (green dot-dashed line). Here, we set $\mu_0=m_\Lambda+m_u+m_d$.}
	\end{center}
\end{figure}

In Fig.~\ref{compare}, taking $\left\langle p_T^2 \right\rangle=1$~GeV$^2$ our theoretical prediction for the $D_c^{\Lambda_c^+}$-FF  at the starting scale $\mu_0$ is shown (dotted  line). In Refs.~\cite{GomshiNobary:1994eq,MoosaviNejad:2016qdx}, it is shown that  the choice of $\left\langle p_T^2 \right\rangle=1$~GeV$^2$ is an optimum value for this quantity  so that any higher value of this parameter will produce the peak position even at lower-z regions.
To check the validity of the Suzuki model, using the parameters presented in Table.~\ref{tab1} we have also plotted the $D_c^{\Lambda_c^+}(z, \mu_0)$-FF at  LO (dashed line), NLO (solid line) and NNLO (dot-dashed line). As is seen, there is a  considerable consistency between both approaches. This allows one to rely on the Suzuki model to determine the heavy quark FFs. As was mentioned previously,   the Suzuki model is much suitable to consider the spin effects of produced hadrons or fragmenting partons which is absent in the phenomenological approach.  It does also enable us to describe the gluon hadronization process which is not well determined in the phenomenological approach.  It also gives one a detailed insight  on the hadronization  process because includes most of the kinematical and dynamical aspects of hadroproduction  process.

%
\section{ $\Lambda_c^+$ Baryon production by top quark decay} \label{sec:B-meson-LHC}

In this section, as a topical application of our baryon FFs, we study the  inclusive single production of  $\Lambda_c^+$  at the CERN LHC. Generally, the $\Lambda_c^+$ baryon may be produced directly or via the decay process of heavier particles,
including  the Higgs boson, the $Z$ boson, and the top quark. At the LHC, the study of energy distributions  of observed hadrons  through top quark decays might be considered as an indirect channel to search for the properties of top quarks.
Since the top quark discovery  by the $D0$ and CDF experiments at Fermilab Tevatron~\cite{Group:2009ad}, the full determination of its properties has not yet been performed so it has been one of the main aims in top physics theories.

In the Standard Model of particle physics, the top has a very short life time ($\tau_t \approx 0.5 \times 10^{-24}$ s~\cite{Chetyrkin:1999ju}) which is much shorter than the typical time to form the QCD bound states, i.e., $\tau_{\rm QCD} \approx 1/\Lambda_{\rm QCD} \approx 3 \times 10^{-24}$ s, then the top quark decays rapidly before hadronization takes place. Related to the Cabibbo-Kobayashi-Maskawa (CKM) mixing matrix element for which $V_{tb} \approx 1$~\cite{Cabibbo:1963yz}, top quarks almost exclusively decay to bottom quarks via $t \to b W^+$ so, in the following, produced bottom quarks hadronize by producing final jets. Therefore, a suggestion  for a new channel to look for top properties at the LHC is to study the inclusive single
$\Lambda_c^+$-baryon production  through the following process
\begin{eqnarray}\label{pros}
	t \rightarrow b W^+ (+g)
	\rightarrow  \Lambda_c^+ W^++ X,
\end{eqnarray}
where, $X$ collectively represents any other final-state particles. 
At the parton level, both the bottom quark and the gluon may hadronize into the $\Lambda_c^+$-baryon so that the gluon fragmentation contributes to the real radiations at NLO. \\
According to the factorization theorem of QCD-improved parton model~\cite{jc}, the partial width of the decay process (\ref{pros})  differential in the scaled
$\Lambda_c^+$-baryon energy, $x_\Lambda$, is expressed as   
\begin{eqnarray}
	\label{eq:master}
	\frac{d\Gamma}{dx_\Lambda}
	=\sum_{i=b,g}
	\int_{x_i^{min}}^{x_i^{max}}
	\frac{dx_i}{x_i}\,
	\frac{d\Gamma_i}{dx_i}
	(\mu_R,\mu_F)
	D_i^{\Lambda_c^+}
	\left(\frac{x_\Lambda}{x_i},
	\mu_F\right),
\end{eqnarray}
where the factorization and the renormalization  scales, i.e.,  $\mu_F$ and $\mu_R$, are arbitrary but to avoid  large logarithms   appearing in the parton differential decay rates $d\Gamma_i/dx_i$,  we set $\mu_R = \mu_F = m_t$, as usual.
For simplicity, we shall work in the top-rest frame in which  the  scaling variables are defined as $x_\Lambda = E_\Lambda/E_b^{\mathrm{max}}$ and
$x_i = E_i/E_b^{\mathrm{max}}$, where $E_\Lambda$ and $E_i$ stand for  the energies of  $\Lambda_c^+$ baryon and parton $i$, respectively. Here, $E_b^{\mathrm{max}}=m_t(1-\omega)/2$ is the maximum energy of the
bottom quark in the process (\ref{pros}), where $\omega=(m_W/m_t)^2$. 
At present, analytic expressions for the Wilson coefficient functions $d\Gamma_i/dx_i$ are
only available at NLO accuracy  which are computed in Refs.~\cite{Nejad:2013fba,Kniehl:2012mn}.
Using our extracted $(b, g) \to \Lambda_c^+$ FFs at LO and NLO, we make our predictions for the energy spectrum of 
$\Lambda_c^+$-baryon produced through the unpolarized top quark decay.
However, a consistent analysis is  presently restricted to NLO approximation, but we also employ the NNLO set of $D_c^{\Lambda_c^+}$-FF  to probe the possible size of NNLO corrections.

\begin{figure}
	\begin{center}
		\includegraphics[width=0.7\linewidth,bb=137 42 690 690]{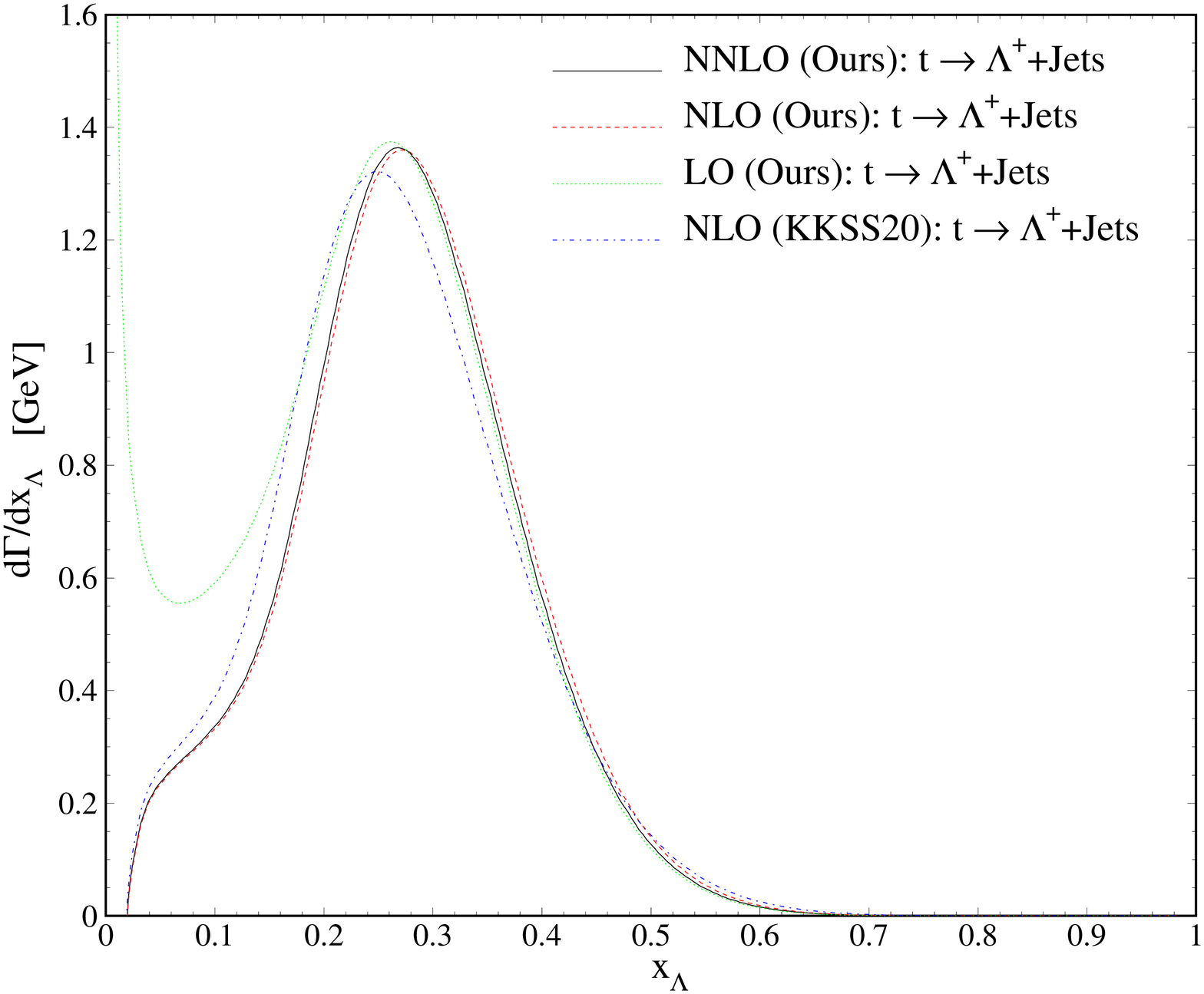}
		\caption{\label{knil}%
			The LO (green dotted line), NLO (red dashed line) and NNLO (black solid line) predictions of $d\Gamma(t\to \Lambda_c^++Jets)/dx_\Lambda$  evaluated with our $\Lambda_c^+$-baryon FF sets. For comparison, the evaluation with the NLO {\tt KKSS20} $\Lambda_c^+$-FF set \cite{Kniehl:2006mw} is also included (blue dot-dashed line). Here, we set  the scale as $\mu=m_t$. }
	\end{center}
\end{figure}

Adopting the input parameters as $m_t=173$ GeV and $m_W=80.379$ GeV, in  Fig.~\ref{knil} we studied the energy distribution of $\Lambda_c^+$-baryon in unpolarized top decays at LO (dotted line),  NLO (dashed line) and NNLO (solid line). 
As is seen, switching from the LO $\Lambda_c^+$-baryon  FF set to  the NLO one slightly smoothens the theoretical prediction, decreasing it in the peak region and the tail region thereunder. In Fig.~\ref{knil}, the results for $d\Gamma/dx_\Lambda$  are also compared to the evaluation with the {\tt KKSS20} $\Lambda_c^+$-baryon  FF set \cite{Kniehl:2020szu}. As is seen, there is a good consistency between both results.
In comparison with the {\tt KKSS20} spectrum, the peak position of our results is shifted towards larger  values of $x_\Lambda$. 

At the LHC, the study of energy distribution of $\Lambda_c^+$-baryon may be also considered as a new window towards searches on new physics. Practically, for the energy distribution of produced hadrons any considerable deviation from the SM predictions  can be assigned to the new physics. For example, it would be a signal for the existence of charged Higgs bosons produced from $t \to \Lambda_c^+H^+$ in the theories beyond the SM \cite{MoosaviNejad:2012ju,MoosaviNejad:2019agw,MoosaviNejad:2016aad,MoosaviNejad:2016jpc,MoosaviNejad:2011yp}. Meanwhile, the study of  $x_\Lambda$-distribution in the decay mode (\ref{pros}) will enable us to deepen our understanding of the nonperturbative aspects of baryon formation by hadronization. Moreover, through studying these distributions the $b/g\to \Lambda_c^+$ FFs can be also constrained event further.

\section{Summary and Conclusions} \label{sec:conclusion}

Through this work, we determined the nonperturbative unpolarized FFs for the charmed baryon $\Lambda_c^+$ in two various approaches; phenomenological analysis and theoretical approach based on the Suzuki model.  Employing the theoretical model we computed the $\Lambda_c^+$-FF at lowest-order of perturbative QCD  whereas using the phenomenological approach we determined the $\Lambda_c^+$-FFs both at LO,  NLO
and, for the first time, at NNLO accuracy in the ZM-VFN scheme by fitting to all available data of
inclusive single $\Lambda_c^+$-baryon production in $e^+e^-$ annihilation from {\tt OPAL} and {\tt Belle} Collaborations \cite{Ackerstaff:1997ki,Niiyama:2017wpp}. A comparison between both approaches showed a good consistency between results.
Note that, the theoretical framework provided by the ZM-VFN scheme is quite appropriate for our data analysis because the characteristic energy scales of annihilation process, i.e., $M_Z$, greatly exceeds the c- and b-quark masses, which could thus be neglected. For our data analysis, we  adopted the same functional form for the parameterization of charm and bottom FFs with three free parameters,  see Eqs.~(\ref{FF parameters}).  From Fig.~\ref{fig2}, it is seen that in the lowest-order approximation the behavior of  theoretical cross section is not acceptable at low-$x_\Lambda$ region  but it is reasonably improved when higher order radiative corrections are considered.  Through  the following aspects, our analysis on  the $\Lambda_c^+$-baryon FFs improves a similar analysis in previous works 
\cite{Kniehl:2006mw,Kniehl:2020szu}. Firstly, we increased the precision of calculation to NNLO, however due to few numbers of experimental data for $\Lambda_c^+$-baryon production our results showed that the effect of this correction is not considerable. Secondly, we did an accurate  estimation of the experimental uncertainties in the
$\Lambda_c^+$-FFs using the Hessian approach. The uncertainties bands of FFs as well as corresponding observables show that the NNLO radiative corrections affect the error band and decrease them considerably. Meanwhile, we have compared, for the first time, the analytical result obtained for the $D_c^{\Lambda_c^+}(z, \mu_0)$-FF through  the Suzuki model with the one extracted via the phenomenological analysis. A good consistency between both approaches ensure the Suzuki model, see Fig.~\ref{compare}. This model is suitable to consider the spin effect of produced hadron or/and initial parton on the corresponding FFs, a subject absent in data analysis approach.  On the other hand, as is well-known,  the gluon FF's play a
significant role in hadroproduction but they  are only feebly constrained by $e^+e^-$ data. But, through the Suzuki model it would be possible to determine them analytically, see Refs.~\cite{MoosaviNejad:2019,Nejad:2014iba,Nejad:2015far}. \\
As a topical application of our obtained FFs, we used the LO,  NLO and NNLO FFs to make our theoretical predictions for the scaled-energy distributions of $\Lambda_c^+$-baryon inclusively produced in unpolarized top decays. This channel is proposed for  independent determination of 
$\Lambda_c^+$-baryon FFs which provides a unique chance to test their universality and
DGLAP scaling violations; two important pillars of the QCD-improved parton model. Furthermore, this study provides a new window towards searches on new physics. \\
For theoretical approach, one can think of other possible improvements  including the Fermi motion of constituents. This is done by considering the real aspects of the valence wave function of baryon  \cite{Brodsky:1985cr,MoosaviNejad:2020svj}, etc. 
Related to the phenomenological approaches, improvements due to the inclusion of finite quark masses and the resummation of soft-gluon logarithms would be effective. 
These effects extend the validity of analysis towards small and large values of $x_\Lambda$, respectively. In this regards, the general-mass variable-flavor-number scheme  (ZM-VFNS) \cite{Nejad:2016epx,Abbaspour:2017pyp} where the charm- and bottom-quark masses are preserved from the beginning 
 provides a
consistent and natural finite mass generalization of the ZM-VFNS on the basis of the $\overline{\mathrm{MS}}$ factorization scheme \cite{Collins:1998rz}.
The implementation of these improvements reaches beyond the scope of our present analysis and is left for future researches.


\end{document}